\documentstyle[12pt,epsfig,axodraw]{article}
\advance\textheight by \topskip
\def\pr#1 #2 #3 { {\rm Phys. Rev.}            {\bf #1}   (#2) #3}
\def\prd#1 #2 #3{ {\rm Phys. Rev.}            {\bf D#1}  (#2) #3}
\def\prl#1 #2 #3{ {\rm Phys. Rev. Lett.}      {\bf #1}   (#2) #3}
\def\plb#1 #2 #3{ {\rm Phys. Lett.}           {\bf B#1}  (#2) #3}
\def\npb#1 #2 #3{ {\rm Nucl. Phys.}           {\bf B#1}  (#2) #3}
\def\prp#1 #2 #3{ {\rm Phys. Rep.}            {\bf #1}   (#2) #3}
\def\zpc#1 #2 #3{ {\rm Z. Phys.}              {\bf C#1}  (#2) #3}
\def\epjc#1 #2 #3{ {\rm Eur. Phys. J.}        {\bf C#1}  (#2) #3}
\def\mpl#1 #2 #3{ {\rm Mod. Phys. Lett.}      {\bf A#1}  (#2) #3}
\def\ijmp#1 #2 #3{{\rm Int. J. Mod. Phys.}    {\bf A#1}  (#2) #3}
\def\ptp#1 #2 #3{ {\rm Prog. Theor. Phys.}    {\bf #1}   (#2) #3}
\def\jhep#1 #2 #3{ {\rm J. High Energy Phys.} {\bf #1}   (#2) #3}
\def\jphg#1 #2 #3{ {\rm J. Phys.}             {\bf G#1}  (#2) #3}
\def\cpc#1 #2 #3{ {\rm Comput. Phys. Commun.} {\bf #1}   (#2) #3}

\newcommand{\be}{\begin{equation}}
\newcommand{\ee}{\end{equation}}
\newcommand{\br}{\begin{eqnarray}}
\newcommand{\er}{\end{eqnarray}}
\newcommand{\ba}{\begin{array}}
\newcommand{\ea}{\end{array}}
\newcommand{\bi}{\begin{itemize}}
\newcommand{\ei}{\end{itemize}}
\newcommand{\bn}{\begin{enumerate}}
\newcommand{\en}{\end{enumerate}}
\newcommand{\bc}{\begin{center}}
\newcommand{\ec}{\end{center}}

\newcommand{\Dir}{\kern -6.4pt\Big{/}}
\newcommand{\Dirin}{\kern -10.4pt\Big{/}\kern 4.4pt}
\newcommand{\DDir}{\kern -8.0pt\Big{/}}
\newcommand{\DGir}{\kern -6.0pt\Big{/}}

\def\frac#1#2{ {{#1} \over {#2} }}

\def\beq{\begin{equation}}
\def\beeq{\begin{eqnarray}}
\def\eeq{\end{equation}}
\def\eeeq{\end{eqnarray}}
\def\a0{\bar\alpha_0}

\def\b0{\beta_0}

\def\ee{e^+e^-}

\def\lms{\Lambda^{(n_{\rm f}=4)}_{\overline{\mathrm{MS}}}}

\def\slashchar#1{\setbox0=\hbox{$#1$}           
     \dimen0=\wd0                                 
     \setbox1=\hbox{/} \dimen1=\wd1               
     \ifdim\dimen0>\dimen1                        
        \rlap{\hbox to \dimen0{\hfil/\hfil}}      
        #1                                        
     \else                                        
        \rlap{\hbox to \dimen1{\hfil$#1$\hfil}}   
        /                                         
     \fi}                                         %

\def\be{\begin{equation}}
\def\ee{\end{equation}}
\def\bea{\begin{eqnarray}}
\def\eea{\end{eqnarray}}

\def\slash{/\kern -5pt}

\def\ims #1 {\ensuremath{M2_{[#1]}}}

\def\s22w{s_{2W}^2}

\begin{document}

\begin{flushright}
{SHEP-05-07}\\ 
{\today}
\end{flushright}

\vspace*{2.0truecm}
\begin{center}
{\Large \bf
Weak effects in proton beam asymmetries\\[0.15 cm] at polarised RHIC and beyond}
\\[1.5cm]
{\large S. Moretti, M.R. Nolten and D.A. Ross}\\[0.15 cm]
{\it School of Physics and Astronomy, University of Southampton}\\
{\it Highfield, Southampton SO17 1BJ, UK}\\[0.25cm]
\end{center}
\vspace*{2.0cm}
\begin{abstract}
\noindent
We report on a calculation of the full one-loop weak corrections through the 
order $\alpha_{\mathrm{S}}^2\alpha_{\mathrm{W}}$ 
to parton-parton scattering in all possible channels at the
Relativistic Heavy Ion Collider (RHIC) running with polarised 
$pp$ beams (RHIC-Spin). This study extends the analysis previously carried 
out for
the case of $2\to2$
subprocesses with two external gluons, by including all
possible four-quark modes with and without an
external gluon. The additional contributions
due to the new four-quarks processes are extremely large, of order 50 to
$100\%$ (of either sign), not only
in the case of parity-violating beam asymmetries but also for
the parity-conserving ones and (although 
to a more limited extent) the total cross section. Such
 ${\cal O}(\alpha_{\mathrm{S}}^2\alpha_{\mathrm{W}})$ effects 
on the CP-violating observables would be
an astounding 5 times larger for the case of the LHC with polarised beams
-- which has been discussed
 as one of the possible upgrades of the CERN machine  -- whereas
they would be much reduced for the case of the CP-conserving ones as
well as the cross section. 
\end{abstract}
\vspace*{1.0cm}
\centerline{Keywords:~{Electroweak physics, Parity violation, Polarised $pp$
scattering, RHIC}}

\newpage

The purely weak component of Electro-Weak (EW) interactions is responsible
for inducing parity-violating effects in jet observables,
detectable through asymmetries in the cross section, which are often regarded 
as an indication of physics beyond the Standard Model
(SM) \cite{reviews}. 
These effects are further enhanced if polarisation
of the incoming beams is exploited, like at the BNL machine
mentioned in the abstract \cite{Bourrely:1990pz,Ellis:2001ba}.
There have also been some discussions \cite{Virey:2001sz,Bass:2001st}
on the idea of polarising the proton beams at the Large Hadron Collider
(LHC) as one of the possible upgrades of the CERN machine, though
no proposal has been put forward yet. At either machine, 
comparison of theoretical predictions 
involving parity-violation with experimental data 
can be used as a powerful tool for confirming or 
disproving  the existence of some beyond the SM scenarios, such as those 
involving right-handed weak currents \cite{Taxil:1997kj}, contact interactions
\cite{Taxil:1996vf} and/or new massive gauge bosons 
\cite{Taxil:1998ni,Taxil:1996vs,Dittmar:2003ir}.

In view of all this,  it becomes of crucial importance to assess
the quantitative relevance of weak effects
entering via  ${\cal O}(\alpha_{\mathrm{S}}^2\alpha_{\mathrm{W}})$
the fifteen possible $2\to2$ partonic
subprocesses responsible for jet production in hadronic 
collisions\footnote{Note that in our treatment we 
identify the jets with the partons
from which they originate.},
namely:
\begin{eqnarray}
g g &\to& q \bar q\\
q \bar q &\to& g g\\
q g &\to& q g\\
\bar q g &\to& \bar q g\\
q q &\to& q q\\
\bar q \bar q &\to& \bar q \bar q\\
q Q &\to& q Q ~({\rm{same~generation}})\\
\bar q \bar Q &\to& \bar q \bar Q ~({\rm{same~generation}})\\
q Q &\to& q Q ~({\rm{different~generation}})\\
\bar q \bar Q &\to& \bar q \bar Q ~({\rm{different~generation}})\\
q \bar q &\to& q \bar q \\
q \bar q &\to& Q \bar Q ~({\rm{same~generation}})\\
q \bar q &\to& Q \bar Q ~({\rm{different~generation}})\\
q \bar Q &\to& q \bar Q ~({\rm{same~generation}})\\
q \bar Q &\to& q \bar Q ~({\rm{different~generation}}),
\end{eqnarray}
with $q$ and $Q$ referring to quarks of different flavours,
limited to $u$-, $d$-, $s$-, $c$- 
and $b$-type (all taken as massless). While the first four
processes (with external gluons) were already computed 
in Ref.~\cite{Ellis:2001ba}, 
the eleven four-quark processes are new to this study\footnote{Note that
$gg\to gg$ does not appear through
${\cal O}(\alpha_{\mathrm{S}}^2\alpha_{\mathrm{W}})$ nor do
$qq'\to QQ'$, $\bar q\bar q'\to \bar Q\bar Q'$ and  $q\bar q'\to Q\bar Q'$.}.
Besides, these four-quark processes
can be (soft and collinear) infrared divergent,
so that gluon bremsstrahlung effects ought to be evaluated to obtain
a finite cross section at the considered order. In addition,
for completeness, we have  also included 
the non-divergent $2\to3$ subprocesses 
\begin{eqnarray}
q g &\to& q q \bar q  \\
\bar q g &\to& \bar q \bar q q\\
q g &\to& q Q \bar Q ~({\rm{same~generation}})\\
\bar q g &\to& \bar q \bar Q Q~({\rm{same~generation}}).
\end{eqnarray}

By recalling that at the typical RHIC-Spin energies (e.g.,
$\sqrt s=300$ and $600$ GeV) the quark luminosity is much larger
than the gluon one, it is clear that are processes with oncoming
quarks 
that dominate the phenomenology of jet production here. In contrast, at the LHC
($\sqrt s=14$ TeV), gluon-induced processes are largely dominant, particularly at low Bjorken$-x$. 
As for what concerns the processes with external gluons, 
it is worth
noticing that no CP-violation occurs at tree-level, so that 
 ${\cal O}(\alpha_{\mathrm{S}}^2\alpha_{\mathrm{W}})$ is the first
non-trivial order at which parity violation
 is manifest. Regarding four-quark processes, 
the following should be noted. Parity-violating 
contributions to channels (5)--(15) are induced already at 
tree-level, through ${\cal O}(\alpha_{\mathrm{EW}}^2)$. Besides, 
all four-quark channels also exist through the CP-violating
${\cal O}(\alpha_{\mathrm{S}}\alpha_{\mathrm{EW}})$
\cite{bgs}, although 
subprocesses (9), (10), (13) and (15) only receive Cabibbo-Kobayashi-Maskawa
(CKM) suppressed  contributions at this accuracy (i.e., 
they mainly proceed via CP-conserving ${\cal O}(\alpha_{\mathrm{S}}^2)$ 
interactions). Furthermore, notice in general that through 
 ${\cal O}(\alpha_{\mathrm{S}}^2\alpha_{\mathrm{W}})$ there are many 
more diagrams available for channels (1)--(15) than via 
${\cal O}(\alpha_{\mathrm{S}}^2)$  or indeed 
${\cal O}(\alpha_{\mathrm{S}}\alpha_{\mathrm{EW}})$ and
${\cal O}(\alpha_{\mathrm{EW}}^2)$. 
Therefore, in terms of parton luminosity, 
simple combinatorics and power counting, 
one should expect the impact of 
${\cal O}(\alpha_{\mathrm{S}}^2\alpha_{\mathrm{W}})$ terms to be large,
certainly in parity-violating observables and possibly
in parity-conserving ones as well. This is what we set out to test in this
paper\footnote{Subprocesses (16)--(19) turn out to be numerically negligible at both
machines and whichever the observable, so that we will not consider them in the remainder.}
for the case of RHIC and LHC. (See Refs.~\cite{Tevatron,Maina:2003is} for an account
of these effects at Tevatron.)

Before proceeding further we ought to clarify at this stage that
we have only computed purely weak effects at one-loop level through
${\cal O}(\alpha_{\mathrm{S}}^2\alpha_{\mathrm{W}})$, while
in the case of tree-level processes 
via ${\cal O}(\alpha_{\mathrm{S}}\alpha_{\mathrm{EW}})$
and ${\cal O}(\alpha_{\mathrm{EW}}^2)$ also the Electro-Magnetic
(EM) contributions are included (and so are the interference effects
between the two). This is why we are referring in this paper to 
the purely weak terms
by adopting the symbol $\alpha_{\mathrm{W}}$, while reserving
the notation 
$\alpha_{\mathrm{EW}}$ for 
the full EW corrections. Here then, we will have
 $\alpha_{\mathrm{W}}\equiv\alpha_{\rm{\small EM}}/\sin^2\theta_{\rm W}$
(with $\alpha_{\rm{\small EM}}$ the Electro-Magnetic (EM) 
coupling constant and $\theta_{\rm W}$ the weak mixing angle) while
 $\alpha_{\mathrm{EW}}$ will refer to the appropriate composition
of QED and weak effects as dictated by the SM dynamics.

We have not computed one-loop EM effects for two reasons. Firstly, their
computation would be 
technically very challenging, because the photon in the loop can become
infrared (i.e., soft and collinear) divergent, thus requiring also
the inclusion of photon bremsstrahlung effects, other than of gluon
radiation. Secondly, ${\cal O}(\alpha_{\mathrm{S}}^2\alpha_{\mathrm{EM}})$
terms (in the above spirit, $\alpha_{\mathrm{EM}}$ signifies here
only the contribution of purely EM interactions)
would carry no parity-violating effects and their contribution
to parity-conserving observables would anyway be overwhelmed by the well
known ${\cal O}(\alpha_{\mathrm{S}}^3)$ terms \cite{Ellis:1985er}
(see also \cite{Ellis:1990ek,Giele:1994gf}). 
However, notice that we are not including these next-to-leading order
(NLO) QCD corrections either, as we are mainly 
interested in parity-violating beam asymmetries.

Since we are considering weak corrections that may be
identified via their induced parity-violating effects and since we wish to
apply our results to the case of polarised proton beams, it is convenient to 
work in terms of helicity Matrix Elements
(MEs). Here, we define the helicity amplitudes by using the formalism
discussed in Ref.~\cite{Maina:2002wz}. 
At one-loop level such helicity amplitudes   
acquire higher order corrections from: 
  (i) self-energy insertions on the fermions and gauge bosons;
 (ii) vertex corrections and
(iii) box diagrams. The expressions for each of the corresponding
one-loop amplitudes
have been calculated using FORM \cite{FORM} and checked by an
independent program based on FeynCalc \cite{FeynCalc}. 
Internal gauge invariance tests have also been performed.
The full expressions for the contributions from these graphs are 
however too lengthy to be reproduced here. 

As already mentioned, infrared divergences  occur when the virtual or 
real (bremsstrahlung) gluon is either soft or collinear 
with the emitting parton
and these have been dealt with by using the formalism
of Ref.~\cite{Catani:1996vz}, whereby corresponding dipole terms 
are subtracted from the bremsstrahlung contributions in order to render the
 phase space integral free of infrared divergences. The
 integrations over the gluon phase space of these dipole terms
 were performed analytically in $d$-dimensions, yielding pole terms
which cancelled explicitly against those of the virtual graphs.
There remains a divergence from the initial state collinear configuration,
which is absorbed into the scale dependence of the Parton Distribution
Functions (PDFs) and must be matched
to the scale at which these PDFs are extracted.
Recall that the remnant initial state collinear divergence at 
${\cal O}(\alpha_{\rm S})$ is absorbed by the LO $Q^2$ dependence of 
the PDFs. Therefore, to ${\cal O}(\alpha_{\rm S}^2 \alpha_{\rm W})$, 
it is sufficient, 
for the purpose of matching these divergences, to consider the LO PDFs. 
It is also consistent to use the values of the running $\alpha_{\rm S}$
obtained form the one-loop $\beta$-function.
In order to display the corrections due to genuine weak interactions the same 
PDFs and strong coupling are used in the LO and 
NLO observables.

The self-energy and vertex correction graphs contain ultraviolet divergences
that have been subtracted here by using the `modified' Dimensional Reduction
(${\overline{\mathrm{DR}}}$) scheme at
the scale $\mu=M_Z$. The use of ${\overline{\mathrm{DR}}}$,
as opposed to the more usual `modified' Minimal Subtraction
(${\overline{\mathrm{MS}}}$) scheme, is forced
upon us by the fact that the $W$- and $Z$-bosons contain axial couplings
which cannot be consistently treated in ordinary dimensional 
regularisation. 
Thus the values taken for the running $\alpha_{\rm S}$ refer to the 
 ${\overline{\mathrm{DR}}}$ scheme whereas the EM coupling,
$\alpha_{\mathrm{EM}}$, 
has been taken to be $1/128$ at the above subtraction
point. The one exception to this
renormalisation scheme has been the case of the self-energy insertions
on external fermion lines, which have been subtracted on mass-shell,
so that the external fermion fields create or destroy particle states
with the correct normalisation

The top quark entering the loops in reactions with external $b$'s has
been assumed to have mass $m_t=175$ GeV
and width $\Gamma_t=1.55$ GeV. The $Z$ mass used was
$M_Z=91.19$ GeV and was related to the $W$-mass, $M_W$, via the
SM formula $M_W=M_Z\cos\theta_W$, where $\sin2\theta_W=0.232$.
(Corresponding widths were $\Gamma_Z=2.5$ GeV and $\Gamma_W=2.08$ GeV.)
For the strong coupling constant, $\alpha_{\rm S}$, we have used the 
one-loop expression
with $\lms$ chosen to match the value required by the 
(LO) PDFs used. The latter were Gehrmann-Stirling 
set A (GSA) \cite{GS} and Gl\"uck-Reya-Stratmann-Vogelsang standard set (GRSV-STN) \cite{GRSV}.

The following beam asymmetries, e.g.,  can be defined, depending on whether
one or both beams are polarised:
\begin{eqnarray}\label{asymmetries}                                \nonumber
A_{LL} \, d\sigma &\equiv &\,d\sigma_{++}\,   - \, d\sigma_{+-} \\ \nonumber 
                  &+      &\,d\sigma_{--}\,   - \, d\sigma_{-+},\\ \nonumber
~~A_{L} \,  d\sigma &\equiv &\,d\sigma_{- }\, ~~- \, d\sigma_{+ },\\ 
A_{PV} \, d\sigma &\equiv &\,d\sigma_{--}\,   - \, d\sigma_{++}.
\end{eqnarray}
The first is parity-conserving while the last two are parity-violating\footnote{In
the numerical analysis which follows we will assume 100\% polarisation of the
beams.}.

\begin{figure}[!t]
\begin{center}
\vspace{-1.cm}
{\epsfig{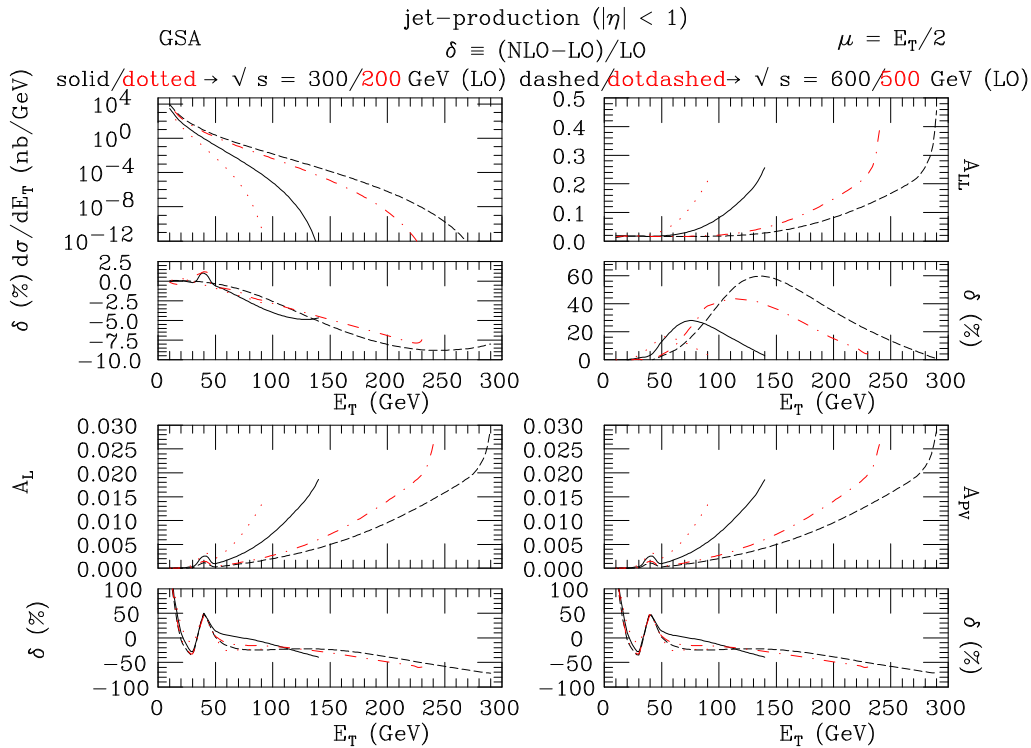}}
\end{center}
\vspace*{-0.75cm}
\caption{\small The dependence of the cross 
section as well as of the beam asymmetries on the jet transverse energy
at tree-level (large frames) and the size of the one-loop weak 
corrections (small frames), at the two RHIC-Spin energies $\sqrt s=300$ 
and 600 GeV. Notice that the pseudo-rapidity range of the jets 
is limited to $|\eta|<1$ and the standard jet cone requirement $\Delta R>0.7$
is imposed as well (although we eventually sum the two- and three-jet contributions).
We use GSA as PDFs and $\mu=E_T/2$ as factorisation/renormalisation scale.
Corresponding results for other two energy options (mentioned later on) $\sqrt s=200$ 
and 500 GeV are also given.}
\label{fig:RHIC}
\end{figure}

Figs.~\ref{fig:RHIC} shows the size of the ${\cal O}(\alpha_{\mathrm{S}}^2\alpha_{\mathrm{W}})$
effects relatively to the well known LO results, the latter being defined as the sum of all
${\cal O}(\alpha_{\mathrm{S}}^2)$,
${\cal O}(\alpha_{\mathrm{S}}\alpha_{\mathrm{EW}})$ and
${\cal O}(\alpha_{\mathrm{EW}}^2)$ contributions, for the case of RHIC,
for two reference energies.  Both the differential cross section and
the above beam asymmetries are plotted, each as a function of the jet transverse energy.
The ${\cal O}(\alpha_{\mathrm{S}}^2\alpha_{\mathrm{W}})$ corrections are already very large
at cross section level, by reaching $-5(-9)\%$  at
$\sqrt s=300(600)$ GeV, in the vicinity of $E_T=120(240)$ GeV. Effects onto the
$A_{LL}$ asymmetry are even larger, with maxima of $\approx 25(60)\%$ for
$E_T\approx70(140)$ GeV, again, in correspondence of $\sqrt s=300(600)$ GeV. In the case
of both $A_{L}$ and $A_{PV}$, in regions away from the threshold at $E_T\approx M_W/4$
(where resonance effects emerge), there is no local maximum for positive or negative corrections,
as both grow monotonically to the level of +100\% (at low $E_T$) and $-50$ to $-70\%$
(at high $E_T$ and with increasing collider energy). All such effects
should comfortably be observable at RHIC, for the customary values  of
integrated luminosity, of 200 and 800 pb$^{-1}$, 
 in correspondence of $\sqrt s=300$ and 600 GeV \cite{reviews}. 

\begin{figure}[!t]
\begin{center}
\vspace{-1.cm}
{\epsfig{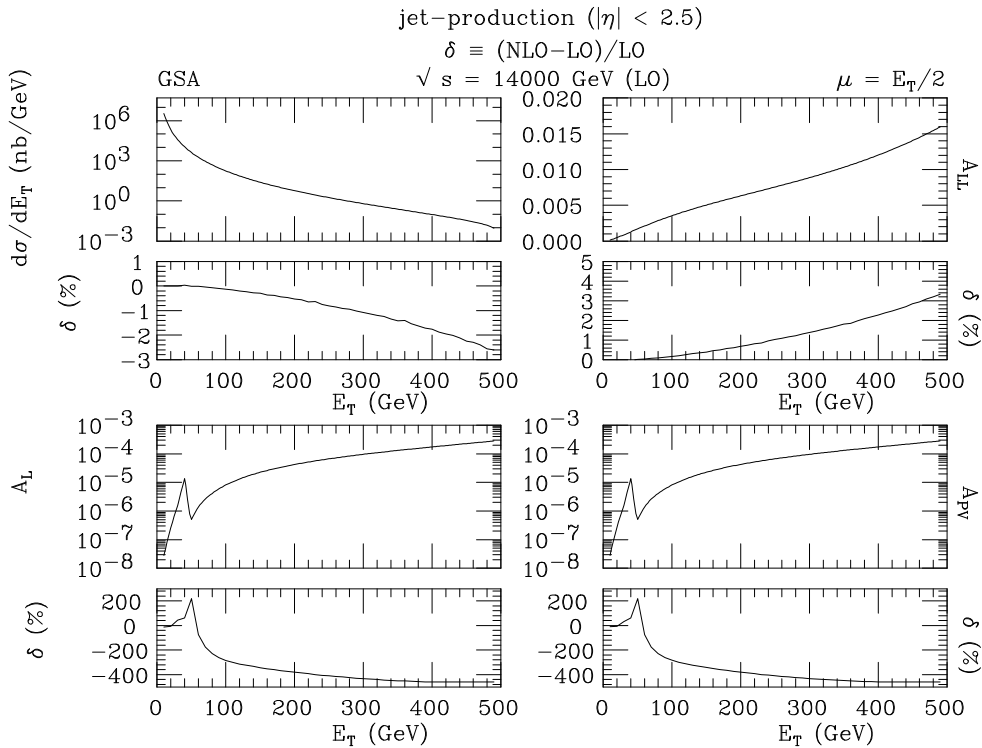}}
\end{center}
\vspace*{-0.75cm}
\caption{\small The dependence of the cross 
section as well as of the beam asymmetries on the jet transverse energy
at tree-level (large frames) and the size of the one-loop weak 
corrections (small frames), at the LHC energy $\sqrt s=14$ 
TeV. Notice that the pseudo-rapidity range of the jets 
is limited to $|\eta|<2.5$ and the standard jet cone requirement $\Delta R>0.7$
is imposed as well (although we eventually sum the two- and three-jet contributions).
We use GSA as PDFs and $\mu=E_T/2$ as factorisation/renormalisation scale.}
\label{fig:LHC}
\end{figure}

At the LHC with polarised beams (but standard energy $\sqrt s=14$ TeV), 
the ${\cal O}(\alpha_{\mathrm{S}}^2\alpha_{\mathrm{W}})$
corrections to the total cross section as
well as the CP-conserving asymmetry are reasonably under control. In fact,
they grow monotonically and reach the $\approx -3\%$ and $\approx 4\%$
at the kinematic limit of the jet transverse energy (as defined by the
PDFs), respectively. However, it is debatable as to whether these effects can actually
be disentangled, as we expect systematic experimental uncertainties to
be of the same order. Away from the threshold at $E_T\approx M_W/2$, 
  ${\cal O}(\alpha_{\mathrm{S}}^2\alpha_{\mathrm{W}})$
effects onto the parity-violating asymmetries are 
instead enormous, as they yield a $K$-factor increasing from $-2$ to $-4.5$, as 
$E_T$ varies from $80$ to 500 GeV. Despite the absolute value of the
CP-violating asymmetries is rather small in the above interval, the
huge LHC luminosity (10 fb$^{-1}$ per year should be feasible for,
say, a 70\% polarisation per beam \cite{AdR}) would render
the above higher order corrections manifest.

It is intriguing to understand the different behaviours of the 
 ${\cal O}(\alpha_{\mathrm{S}}^2\alpha_{\mathrm{W}})$ effects depending
on the observable and the collider being considered. To this end,
we have presented in Tabs.~\ref{tab:RHIC}--\ref{tab:LHC} the 
contributions to the $E_T$ dependent cross section of subprocesses (1)--(15) through
${\cal O}(\alpha_{\mathrm{S}}^2\alpha_{\mathrm{W}})$ 
separately, at both RHIC-Spin and LHC. 
The purpose of these tables is to illustrate that the leading partonic composition of the 
${\cal O}(\alpha_{\mathrm{S}}^2\alpha_{\mathrm{W}})$ corrections is markedly different at
the two machines. While at RHIC the key role is played by subprocess (7), at the LHC 
the conspicuous rise of the gluon-luminosity
enhances in turn the yield of channel (3) to a level
comparable to that of mode (7)\footnote{The relevance of 
the latter throughout originates from the combination
of a always sizable valence quark luminosity and a large Feynman diagram combinatorics,
as opposed to, e.g., a gluon luminosity steeply increasing with the collider energy 
but combined with a small numbers of graphs \cite{Ellis:2001ba}.}.
 (The hierarchy
among the subprocesses seen in Tabs.~\ref{tab:RHIC}--\ref{tab:LHC}, for fixed
jet transverse energy, is characteristic across most of the available $E_T$ range
at both colliders.) The ${\cal O}(\alpha_{\mathrm{S}}^2\alpha_{\mathrm{W}})$ 
corrections are  particularly large for subprocesses (7)--(8) and (12), mainly in virtue
of the large combinatorics involved at loop level (as intimated earlier), with respect
to the LO case.

\begin{table}[h]
    \caption{\small The contributions of subprocesses (1)--(15) to 
 order $\alpha_{\mathrm{S}}^2\alpha_{\mathrm{W}}$ with respect to the
full LO result for the total
cross section at RHIC-Spin, at $E_T= 70$ GeV for $\sqrt s=300$ GeV and 
                               $E_T=140$ GeV for $\sqrt s=600$ GeV.
Here, we have paired
together the channels with identical Feynman diagram topology.
We use GSA as PDFs and $\mu=E_T/2$ as factorisation/renormalisation scale.
Column (a) is the percentage contribution from  the  ${\cal O}(\alpha_{\mathrm{S}}^2\alpha_{\mathrm{W}})$
corrections, column (b) is the percentage correction to the tree-level partonic subprocess
and column (c) is the percentage contribution at the tree-level of that partonic subprocess to the
differential cross section at the relevant $E_T$.}
    \begin{center}
    \begin{tabular}{|l||c|c|c||c|c|c|} \hline
     Subprocess & \multicolumn{3}{c ||}{$\sqrt s=300$  GeV} &
      \multicolumn{3}{c |}{ $\sqrt s=600$  GeV} \\ \hline \hline
 & (a) & (b) & (c) &(a) &(b) &(c) \\ \hline\hline    
        $gg\to gg$     & & & 1.35 & & & 1.17\\ \hline 
               (1)     &1.73E-05 & 0.0267 & 0.0650 &--0.000101 & --0.179 & 0.0565\\ \hline 
               (2)     &4.72E-05 & 0.0253 & 0.188 &--0.000331 & --0.179 & 0.184\\ \hline 
          (3)--(4)     &--0.0144 & --0.0589 & 24.7   &--0.0608 & --0.264 & 23.0\\ \hline 
          (5)--(6)     &--0.411 & --0.899 & 46.1    &--1.42 & --3.00 & 47.1\\ \hline 
          (7)--(8)     &--1.57 & --6.64 & 23.8    &--3.65 & --14.7 & 24.6\\ \hline 
         (9)--(10)     &0.000393 & 0.0551 & 0.718   &--0.00602 & --0.843 & 0.711\\ \hline 
              (11)     &--0.00236 & --0.252 & 0.946    &--0.00664 & --0.714 & 0.927\\ \hline 
              (12)     &0.00664 & 10.90 & 0.0613     &0.0146 & 25.5 & 0.0571\\ \hline 
              (13)     &0.00222 & 1.23 & 0.182      &0.00199 & 1.17 & 0.169\\ \hline 
              (14)     &0.0406 & 3.20 & 1.28      &0.0282 & 2.24 & 1.25\\ \hline 
              (15)     &0.000350 & 0.0491 & 0.717 &--0.00572 & --0.803 & 0.708\\ \hline 
$qq'\to QQ'$ or $\bar q\bar q'\to \bar Q\bar Q'$  & & &0.00710 & & & 0.0148\\ \hline  
$q\bar q'\to Q\bar Q'$                       & & & 0.00234 & & & 0.00131\\ \hline 

          Total        &--1.94 & &                  &--5.11 & & \\ \hline 
\end{tabular}
    \end{center}
\label{tab:RHIC}
\end{table}

\begin{table}[h]
    \caption{\small The contributions of subprocesses (1)--(15) to 
 order $\alpha_{\mathrm{S}}^2\alpha_{\mathrm{W}}$ with respect to the
full LO result for the total
cross section at LHC, at $E_T=300$ GeV for $\sqrt s=14$ TeV.
Here, we have paired
together  the channels with identical Feynman diagram topology.
We use GSA as PDFs and $\mu=E_T/2$ as factorisation/renormalisation scale.
Column (a) is the percentage contribution from  the  ${\cal O}(\alpha_{\mathrm{S}}^2\alpha_{\mathrm{W}})$
corrections, column (b) is the percentage correction to the tree-level partonic subprocess and
and column (c) is the percentage contribution at the tree-level of that partonic subprocess to the
differential cross section at the relevant $E_T$.}
    \begin{center}
    \begin{tabular}{|l||c|c|c|} \hline
     Subprocess & \multicolumn{3}{c |}{$\sqrt s=14$  TeV}  \\ \hline \hline
     & (a) & (b) & (c) \\ \hline \hline    
        $gg\to gg$     & & & 41.9\\ \hline 
               (1)     &--0.0315 & --1.25 & 1.89\\ \hline 
               (2)     &--0.00386 & --1.27 & 0.228\\ \hline 
          (3)--(4)     &--0.455 & --0.711 & 47.8\\ \hline 
          (5)--(6)     &--0.112 & --4.92 & 1.70\\ \hline 
          (7)--(8)     &--0.431 & --17.3 & 1.87\\ \hline 
         (9)--(10)     &--0.0330 & --2.67 & 0.926\\ \hline 
              (11)     &--0.0328 & --1.85 & 1.328\\ \hline 
              (12)     &0.0466 & 64.8 & 0.0540\\ \hline 
              (13)     &--0.00316 & --1.50 & 0.158\\ \hline 
              (14)     &0.0131 & 0.821 & 1.196\\ \hline 
              (15)     &--0.0325 & --2.64 & 0.924\\ \hline 
$qq'\to QQ'$ or $\bar q\bar q'\to \bar Q\bar Q'$  & & & 0.0221\\ \hline  
$q\bar q'\to Q\bar Q'$                       & & & 0.000979\\ \hline 
          Total        &--1.075 & & \\ \hline 
\end{tabular}
    \end{center}
\label{tab:LHC}
\end{table}

The different behaviours seen in Figs.~\ref{fig:RHIC}--\ref{fig:LHC} 
can easily be interpreted in terms of the LO contributions. In this respect, as
already mentioned, Tabs.~\ref{tab:RHIC}--\ref{tab:LHC} clearly make the point
that the jet phenomenology at RHIC-Spin is dominated by subprocesses initiated by
quarks only while at the LHC gluons-induced channels are generally predominant.
At RHIC energies, LO production through order $\alpha_{\mathrm{S}}^2$ is dominated
by channel (5) whereas at the LHC the overwhelmingly dominant $\alpha_{\mathrm{S}}^2$ channels are $gg\to gg$ (which
is not subject to  ${\cal O}(\alpha_{\mathrm{S}}^2\alpha_{\mathrm{W}})$ corrections, as already
mentioned) and subprocess (3). Besides,
channel (5) through ${\cal O}(\alpha_{\mathrm{S}}^2)$ is mainly concentrated
at low $E_T$ while with growing  $E_T$ 
the   ${\cal O}(\alpha_{\mathrm{S}}\alpha_{\mathrm{EW}})$ and -- particularly --
${\cal O}(\alpha_{\mathrm{EW}}^2)$ terms gain in relative importance. Furthermore,  ${\cal O}(\alpha_{\mathrm{S}}^2)$
terms entering channel (5) do not contribute, obviously, to the parity-violating
asymmetries. Therefore, it should not be surprising
to see at RHIC-Spin that our corrections are very large in the case of
the latter, where the LO term is  ${\cal O}(\alpha_{\mathrm{S}}\alpha_{\mathrm{EW}})$,
respect to which the corrections computed here are suppressed only by one power 
of $\alpha_{\mathrm{S}}$.  We attribute instead the size of the 
${\cal O}(\alpha_{\mathrm{S}}^2\alpha_{\mathrm{W}})$ effects on the 
cross section and the parity-conserving asymmetry 
again to the fact that through 
 ${\cal O}(\alpha_{\mathrm{S}}^2\alpha_{\mathrm{W}})$ there are many 
more diagrams available for such channels than via 
${\cal O}(\alpha_{\mathrm{S}}^2)$  or indeed 
${\cal O}(\alpha_{\mathrm{S}}\alpha_{\mathrm{EW}})$ and
${\cal O}(\alpha_{\mathrm{EW}}^2)$.  As for the LHC,
the fact that $gg\to gg$ and  subprocess (3) vastly 
dominates through ${\cal O}(\alpha_{\mathrm{S}}^2)$
the $d\sigma/dE_T$ distribution explains
why  ${\cal O}(\alpha_{\mathrm{S}}^2\alpha_{\mathrm{W}})$ 
corrections are limited to the percent level. In $A_{LL}$, which 
has no $gg\to gg$ component, 
${\cal O}(\alpha_{\mathrm{S}}^2\alpha_{\mathrm{W}})$ effects
become somewhat more visible in comparison. 

Furthermore, in the case of the LHC, 
one should note the monotonic rise of the corrections
with increasing jet transverse energy, for all observables studied,
which can be attributed to  the so-called
Sudakov (leading) logarithms  \cite{Melles:2001ye,Denner:2001mn} 
of the form 
$\alpha_{\mathrm{W}}\log^2({E_T^2}/M_{W}^2)$, which appear
in the presence of higher order weak corrections.
These `double logarithms' are due to a lack of cancellation of infrared 
(both soft and collinear) virtual and real emission in 
higher order contributions due to $W$-exchange, arising from a  
violation of the Bloch-Nordsieck theorem occurring in non-Abelian theories. 
(In fact, if events with real $Z$ radiation are
vetoed in the jet sample, $\alpha_{\mathrm{W}}\log^2(E_T/M_{Z}^2)$ terms
would also affect the corrections \cite{Tevatron}.) Clearly, at LHC energies, $E_T$
can be very large, thus probing the kinematic regime of
 these logarithmic effects, which 
instead affected RHIC only very mildly. 
Combine then the effects of such large logarithms with
the fact that $A_{L}$ and $A_{PV}$ receive no pure QCD
contributions, and one can explain the enormous (and increasing
with $E_T$) ${\cal O}(\alpha_{\mathrm{S}}^2\alpha_{\mathrm{W}})$  corrections
to these two observables. 
 In fact,  recall that another way of viewing the
 ${\cal O}(\alpha_{\mathrm{S}}^2\alpha_{\mathrm{W}})$  terms computed here
is as first order QCD corrections to the
${\cal O}(\alpha_{\mathrm{S}}\alpha_{\mathrm{W}})$ terms,
which are  the leading order contributions to  $A_{L}$ and $A_{PV}$. 
From this perspective then, the large results 
reported here can be understood as large ${\cal O}(\alpha_{\rm S})$ 
corrections.
Furthermore, also recall here the following two aspects, already mentioned.
Firstly, there are several partonic processes which are  CKM suppressed
at  ${\cal O}(\alpha_{\mathrm{S}} \alpha_{\rm EW})$ but which occur without
CKM suppression at  ${\cal O}(\alpha_{\mathrm{S}}^2 \alpha_{\rm W})$.
Secondly, processes involving external gluons and weak interactions
occur for the first time at  ${\cal O}(\alpha_{\mathrm{S}}^2 \alpha_{\rm W})$.

\begin{figure}[!t]
\begin{center}
\vspace{-1.cm}
{\epsfig{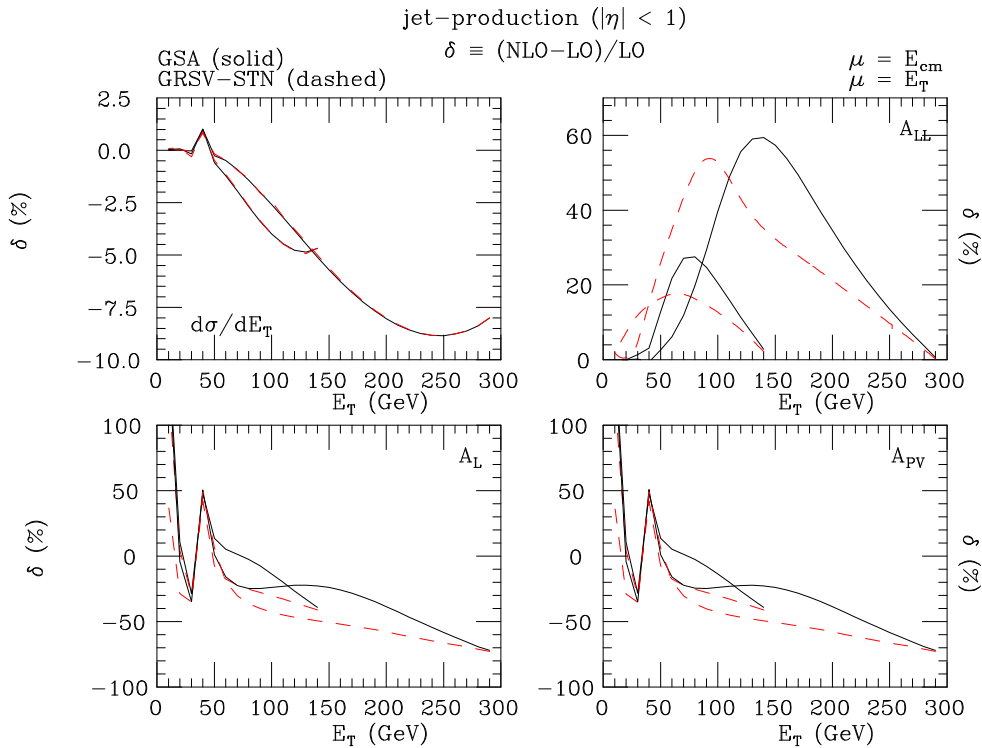}}
\end{center}
\vspace*{-0.75cm}
\caption{\small The dependence of the corrections to the cross 
section as well as the beam asymmetries on the jet transverse energy
for two sets of PDFs, GSA and GRSV-STN, at the two RHIC-Spin energies $\sqrt s=300$
(curves extending to 150 GeV) 
and 600 GeV (curves extending to 300 GeV). Notice that the pseudo-rapidity range of the jets 
is limited to $|\eta|<1$ and the standard jet cone requirement $\Delta R>0.7$
is imposed as well (although we eventually sum the two- and three-jet contributions).
We use $\mu=E_{\rm {cm}}(E_T)$ as factorisation/renormalisation scale in the case of  GSA(GRSV-STN).}
\label{fig:RHIC_NewPDFs}
\end{figure}

\begin{figure}[!t]
\begin{center}
\vspace{-1.cm}
{\epsfig{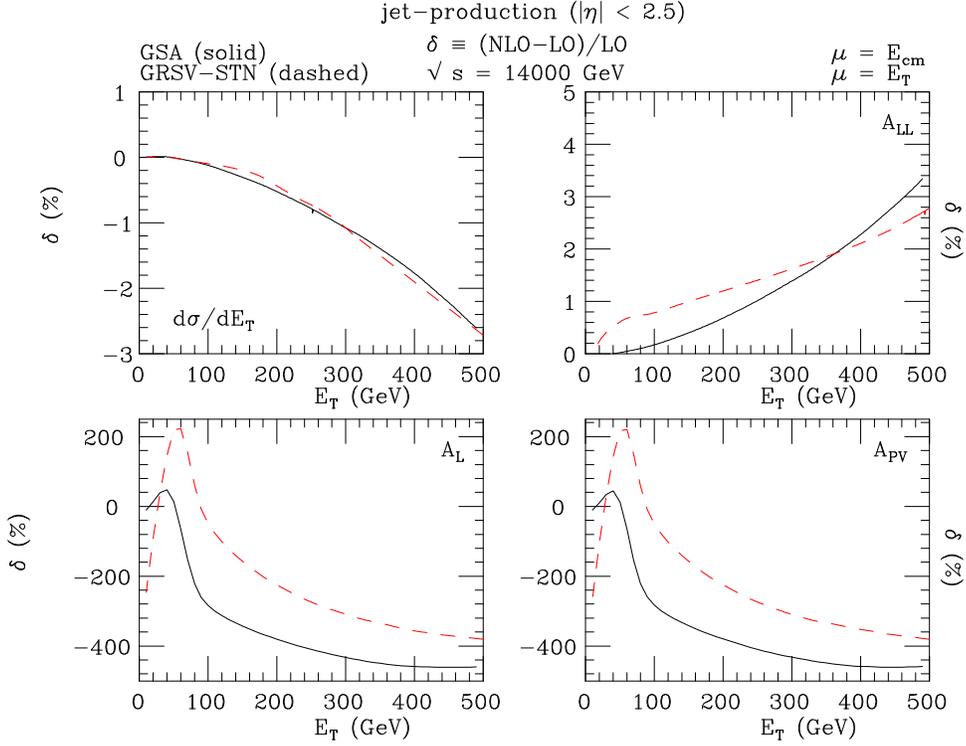}}
\end{center}
\vspace*{-0.75cm}
\caption{\small The dependence of the corrections to the cross 
section as well as the beam asymmetries on the jet transverse energy
for two sets of PDFs, GSA and GRSV-STN, at the LHC energy $\sqrt s=14$ 
TeV.  Notice that the pseudo-rapidity range of the jets 
is limited to $|\eta|<2.5$ and the standard jet cone requirement $\Delta R>0.7$
is imposed as well (although we eventually sum the two- and three-jet contributions).
We use $\mu=E_{\rm {cm}}(E_T)$ as factorisation/renormalisation scale in the case of  GSA(GRSV-STN).}
\label{fig:LHC_NewPDFs}
\end{figure}

As one of the purposes of polarised colliders is to measure polarised
structure functions, in the ultimate attempt to reconstruct the proton
spin, it is of some relevance to see how the ${\cal O}(\alpha_{\mathrm{S}}^2\alpha_{\mathrm{W}})$  
results obtained so far for GSA compare against GRSV-STN.   This is
done in Figs.~\ref{fig:RHIC_NewPDFs}--\ref{fig:LHC_NewPDFs}, where we
have also adopted the different choice $\mu=E_{\rm{cm}}(E_T)$
as factorisation/renormalisation scale, i.e., the centre-of-mass
energy at parton level $\sqrt{\hat s}$(jet transverse energy), 
for the GSA(GRSV-STN) set. A comparison between the GSA curves in  
Figs.~\ref{fig:RHIC_NewPDFs}--\ref{fig:LHC_NewPDFs} and those in
Figs.~\ref{fig:RHIC}--\ref{fig:LHC} reveals that the scale
dependence of our corrections is not very substantial for a given PDF
set (the same is true for the case of GRSV-STN). In contrast,
depending on the choice of PDFs, corrections through  
 ${\cal O}(\alpha_{\mathrm{S}}^2\alpha_{\mathrm{W}})$
can be very different for each observables studied at
both RHIC-Spin and LHC, with the exception of the
cross section in either case.

Altogether, the results presented here point to the extreme relevance of one-loop
${\cal O}(\alpha_{\rm{S}}\alpha_{\rm{W}}^2)$ weak contributions for precision
analyses of jet data produced in polarised proton-proton scattering at RHIC.
We have confirmed that this would be the case also at a polarised
LHC, which has been discussed as one of the possible upgrades of
the CERN collider. The size of the afore-mentioned
corrections, relative to the lowest order results, 
is rather insensitive to the choice of factorisation/renormalisation scale,
yet it shows some sizable dependence on the polarised PDFs used. 
EM effects were neglected here
because they are not subject to  parity-violating effects. 
However, their computation is currently in progress. 
 The inclusion of NLO terms from pure QCD,
 through ${\cal O}(\alpha_{\rm{S}}^3)$, 
is also in order, as they can produce effects of order 100\%, 
 even to the parity-violating asymmetries, though in this case they will 
only amount
to a rescaling (within a factor of 2 at the most) of the normalisation, not
to a change in shape. We are now working towards the full  ${\cal O}(\alpha_{\rm{S}}^3)$ 
results including beam polarisation effects \cite{preparation}.

 Finally, extrapolation of our results
to other collider energies, chiefly for RHIC,
for operation  at $\sqrt s=200$ and 500 GeV,
is straightforward. As expected, these
 results do not differ dramatically from those
obtained  for 300 and 600 GeV, respectively. Fig.~\ref{fig:RHIC} also
shows the size of the corrections at these two additional energies, for
our usual observables and default choice of PDFs and factorisation/renormalisation
scale. For the purpose
of emulating the effects of ${\cal O}(\alpha_{\rm{S}}\alpha_{\rm{W}}^2)$ 
terms at whichever collider and energy, we make available our code on request.

\section*{Acknowledgements}

We thank Ezio Maina for collaborative work in the early stages of this
analysis.

\end{document}